\documentclass[usegraphicx,usenatbib]{mn2e}

\begin{document}

\title[Origins of the millimeter background]{Origins
of the extragalactic background at 1$\,$mm from a combined
analysis of the AzTEC and MAMBO data in GOODS-N}

\author[Penner et al.]{Kyle Penner,$^{1}$\thanks{kpenner@as.arizona.edu}
Alexandra Pope,$^{2}$\thanks{\emph{Spitzer} Fellow}
Edward L. Chapin,$^{3}$
Thomas R. Greve,$^{4,5}$
\newauthor
Frank Bertoldi,$^{6}$
Mark Brodwin,$^{7}$\thanks{W. M. Keck Postdoctoral Fellow}
Ranga-Ram Chary,$^{8}$
Christopher J. Conselice,$^{9}$
\newauthor
Kristen Coppin,$^{10}$
Mauro Giavalisco,$^{11}$
David H. Hughes,$^{12}$
Rob J. Ivison,$^{13,14}$
\newauthor
Thushara Perera,$^{15}$
Douglas Scott,$^{3}$
Kimberly Scott$^{16}$ and Grant Wilson$^{11}$ \\
$^{1}$Department of Astronomy, University of Arizona, 933 N. Cherry Ave., Tucson, AZ 85721, USA  \\
$^{2}$National Optical Astronomy Observatory, 950 N. Cherry Ave., Tucson AZ 85719, USA \\
$^{3}$Dept. of Physics and Astronomy, University of British Columbia, 6224 Agricultural Road, Vancouver, B.C. V6T 1Z1, Canada \\
$^{4}$Max-Planck Institute for Astronomy, K\"onigstuhl 17, 69117 Heidelberg, Germany \\
$^{5}$Dark Cosmology Centre, Juliane Maries Vej 30, 2100 Copenhagen \O, Denmark \\
$^{6}$Argelander Institute for Astronomy, University of Bonn, Auf dem Huegel 71, 53121 Bonn, Germany \\
$^{7}$Harvard-Smithsonian Center for Astrophysics, 60 Garden Street, Cambridge, MA 02138, USA \\
$^{8}$Spitzer Science Center, California Institute of Technology, Pasadena, CA 91125, USA \\
$^{9}$School of Physics and Astronomy, University of Nottingham, Nottingham NG7 2RD, UK \\
$^{10}$Institute for Computational Cosmology, Durham University, South Road, Durham DH1 3LE, UK \\
$^{11}$Department of Astronomy, University of Massachusetts, Amherst, MA 01003, USA \\
$^{12}$Instituto Nacional de Astrof\'{i}sica, \'{O}ptica y Electr\'{o}nica, Luis Enrique Erro No. 1,
Tonantzintla, Puebla, C.P. 72840, Mexico \\
$^{13}$UK Astronomy Technology Centre, Royal Observatory, Blackford Hill, Edinburgh EH9 3HJ, UK \\
$^{14}$Institute for Astronomy, University of Edinburgh, Blackford Hill, Edinburgh EH9 3HJ, UK \\
$^{15}$Department of Physics, Illinois Wesleyan University, Bloomington, IL 61701, USA \\
$^{16}$Department of Physics and Astronomy, University of Pennsylvania, Philadelphia, PA 19104, USA
}

\maketitle
\begin{abstract}
We present a study of the cosmic infrared background, which is a measure of the
dust obscured activity in all galaxies in the Universe.
We venture to isolate the galaxies responsible
for the background at 1$\,$mm; with spectroscopic and photometric redshifts
we constrain the redshift distribution of these galaxies.  We create a
deep 1.16$\,$mm map ($\sigma \sim 0.5$$\,$mJy) by combining 
the AzTEC 1.1$\,$mm and MAMBO 1.2$\,$mm datasets in GOODS-N.
This combined map contains 41 secure detections, 13 of which are new.
By averaging the 1.16$\,$mm flux densities of individually
undetected galaxies with 24$\,$$\micron$ flux densities $> 25$\,$\mu$Jy,
we resolve 31--45 per cent of the 1.16$\,$mm background.
Repeating our analysis on the SCUBA 850$\,$$\micron$ map,
we resolve a higher percentage (40--64 per cent) of the
850$\,$$\micron$ background.  A majority
of the background resolved (attributed to individual galaxies)
at both wavelengths comes from galaxies at $z > 1.3$.
If the ratio of the resolved
submillimeter to millimeter background is applied to a reasonable scenario
for the origins of the unresolved submillimeter background,
60--88 per cent of the total 1.16$\,$mm background
comes from galaxies at $z > 1.3$.
\end{abstract}

\section{Introduction}

The cosmic infrared background (CIB) is the total dust emission
from all galaxies in the Universe.  The contribution of galaxies to the background
varies with redshift; this variation constrains the evolution over cosmic time
of the output of dust obscured AGN activity and star formation.
Decomposing the background into individual galaxies provides
constraints as a function of redshift on the processes important to galaxy evolution.

Models predict that a large fraction of the CIB
at longer (sub)millimeter wavelengths comes from galaxies at high redshift
\citep*{gispert00}.  The main evidence is that the SED of the (sub)mm background
is less steep than the SED of a representative (sub)mm galaxy;
the shallow slope of the background can be due to high redshift galaxies,
so that the peak of their infrared SED shifts to observed (sub)mm wavelengths
\citep{lagache05}.  In this paper we address the question:  `What galaxies are responsible for
the CIB at $\lambda \sim 1$$\,$mm, and what is their redshift distribution?'

It is difficult to individually detect a majority of the galaxies that contribute
to the millimeter background, as maps are limited by confusion noise due to the
large point spread functions of current mm telescopes.
To resolve the $\sim$\,$1$$\,$mm background, we rely
on a stacking analysis of galaxies detected at other wavelengths.
Stacking is the process of averaging the millimeter flux density
of a large sample of
galaxies \emph{not} individually detected in a millimeter map;
the desired result is a high
significance detection of the `external' sample as a whole (or in bins of
flux density, redshift, etc.).

Stacking the (sub)mm flux density of galaxies is not a new methodology.
Several studies seek to decompose the background at 850$\,$$\micron$ by stacking on SCUBA
maps (\citealt{wang06}; \citealt{dye06}; \citealt{serjeant08}).
These studies agree that the 850$\,$$\micron$ background is not completely resolved by
current samples of galaxies; however, they reach contradictory conclusions on the redshift
distribution of the galaxies that contribute to the resolved background.
Recently, stacking has been carried out on BLAST maps at
250, 350, and 500$\,$$\micron$ (\citealt{marsden09};
\citealt{devlin09}; \citealt{pascale09}; \citealt{chary10}).
As with stacking on any map with a large PSF, stacking on BLAST maps is subject to complications
when the galaxies are angularly clustered.  We take this issue into consideration
in our analysis in this paper.

We combine the AzTEC 1.1$\,$mm \citep{perera08} and
MAMBO 1.2$\,$mm \citep{greve08} maps in the
GOODS-N field to create a deeper map at an effective wavelength of 1.16$\,$mm.
A significant advantage of the combined 1.16$\,$mm map over the individual
1.1$\,$mm and 1.2$\,$mm maps is reduced noise.
We investigate the contribution
of galaxies with 24$\,$$\micron$ emission to the 1.16$\,$mm background
as a function of redshift.
By stacking the same sample of galaxies
on the SCUBA 850$\,$$\micron$ map in GOODS-N, we calculate the relative contribution
of galaxies to the background at 850$\,$$\micron$ and 1.16$\,$mm as a function of
redshift; we use this to infer the
redshift distribution of the galaxies contributing to the
remaining, unresolved 1.16$\,$mm background.

This paper is organized as follows: in \S\ref{sec:data}, we describe
the data and our analysis of the data; in \S\ref{sec:stacking} we
describe stacking and several considerations when performing a
stacking analysis.  We present our results in \S\ref{sec:results},
and conclude in \S\ref{sec:conclusions}.

\section{Data}\label{sec:data}

\subsection{Creating the combined 1.16$\,$mm map}

There are two deep millimeter surveys of the Great Observatories Origins Deep Survey
North region \citep[GOODS-N;][]{dickinson03}.  The AzTEC survey
at 1.1$\,$mm carried out on the JCMT (PSF FWHM = 19.5$\,$arcsec)
reaches a $1\sigma$ depth of 0.96 mJy over 0.068
deg$^{2}$ \citep{perera08}. The MAMBO survey at 1.2$\,$mm carried out on the IRAM 30-m
telescope (PSF FWHM = 11.1$\,$arcsec) reaches a $1\sigma$ depth of 0.7 mJy over
0.080 deg$^{2}$ \citep{greve08}.  The noise values refer to the uncertainty in determining
the flux density of a point source.  For more details on the individual maps, we
refer the reader to those papers.

We create a combined mm map from a weighted average of the AzTEC 1.1$\,$mm
and MAMBO 1.2$\,$mm maps.
We use the PSF-convolved maps that are on the same RA and Dec
grid with the same pixel size (2$\,$arcsec$\,$$\times$$\,$2$\,$arcsec).

The weighted
average flux density in a pixel in the combined mm map is calculated as:
\begin{equation}
S_{\mathrm{measured}} = \frac{\frac{w_{\mathrm{A}}S_{\mathrm{A}}}{\sigma_{\mathrm{A}}^{2}} +
          \frac{w_{\mathrm{M}}S_{\mathrm{M}}}{\sigma_{\mathrm{M}}^{2}}}
          {\frac{w_{\mathrm{A}}}{\sigma_{\mathrm{A}}^{2.}}+\frac{w_{\mathrm{M}}}{\sigma_{\mathrm{M}}^{2}}},
\label{eqn:coadd}
\end{equation}
where $S_{\mathrm{A}}$ and $\sigma_{\mathrm{A}}$ are the measured flux density and noise in the
AzTEC 1.1$\,$mm map, $S_{\mathrm{M}}$ and $\sigma_{\mathrm{M}}$ are the measured flux density and noise in the
MAMBO 1.2$\,$mm map, and the $w$'s are constants.

The noise in each pixel from Eq.~\ref{eqn:coadd} is thus
\begin{equation}
\sigma = \frac{\sqrt{\frac{w_{\mathrm{A}}^{2}}{\sigma_{\mathrm{A}}^{2}} +
                     \frac{w_{\mathrm{M}}^{2}}{\sigma_{\mathrm{M}}^{2}}}}
                     {\frac{w_{\mathrm{A}}}{\sigma_{\mathrm{A}}^{2}} +
                      \frac{w_{\mathrm{M}}}{\sigma_{\mathrm{M}}^{2}}}.
\label{eqn:coaddrms}
\end{equation}
Use of the inverse variance weights in combining the two maps
results in the map with minimum noise.  We are instead interested in the
resulting map with the \emph{maximum} signal-to-noise ratio (SNR) of the
sources, whether these sources are above or below some detection threshold.
We introduce additional weights, $w_{\mathrm{A}}$ and $w_{\mathrm{M}}$,
which are constant multiplicative factors for the two individual
maps.  To rephrase the justification for these
$w$'s in astrophysical terms -- at (sub)millimeter
wavelengths, the spectral energy distributions (SEDs) of galaxies
fall off $\propto \nu^{2+\beta}$ (a Rayleigh-Jeans fall off with
emissivity index $\beta$); the flux density at 1.1$\,$mm is higher than
that at 1.2$\,$mm.
A simple inverse variance weighted
average ($w_{\mathrm{A}} = w_{\mathrm{M}}$) does not account for this.

\begin{figure}
\centering
\includegraphics[scale=.44]{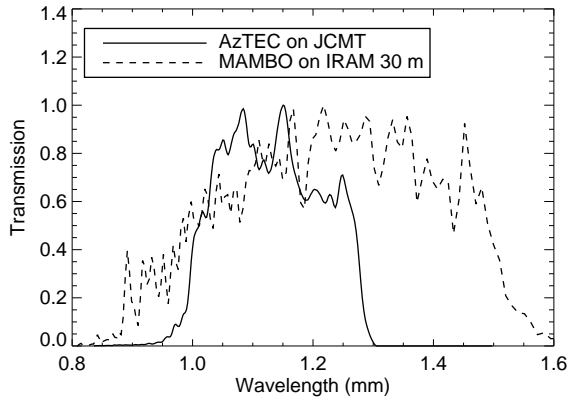}
\caption{Transmission curves for the AzTEC and MAMBO detectors on their respective
telescopes.
\label{fig:trans}}
\end{figure}

The optimal $w$'s come from iteratively maximizing the SNR of the detections
in the resulting combined map (in practice, we maximize the number of
detections above 3.8$\sigma$).  The two values are [$w_{\mathrm{A}}$, $w_{\mathrm{M}}$]
= [0.56, 0.44].  Given these $w$'s, the inverse variance weights, and that
the transmission curves for the individual maps shown in Fig. \ref{fig:trans} overlap,
the central wavelength of the combined map is 1.16$\,$mm.
In the absence of \emph{any} weighting, the combined map has an effective wavelength of
1.15$\,$mm. Weighting the individual maps results in a small shift of the central wavelength
of the combined map to 1.16$\,$mm.

The combined 1.16$\,$mm map has two significant advantages over the individual
1.1$\,$mm and 1.2$\,$mm maps: 1) reduced noise (by roughly $\sqrt{2}$); and 2)
increased reliability of secure detections.
The AzTEC and
MAMBO catalogues include some spurious detections
(\citealt{perera08}; \citealt{greve08}); by combining
the two (independent) maps, the secure detections in the resulting map
may be more reliable (this is the expectation).

\begin{figure}
\centering
\includegraphics[scale=.44]{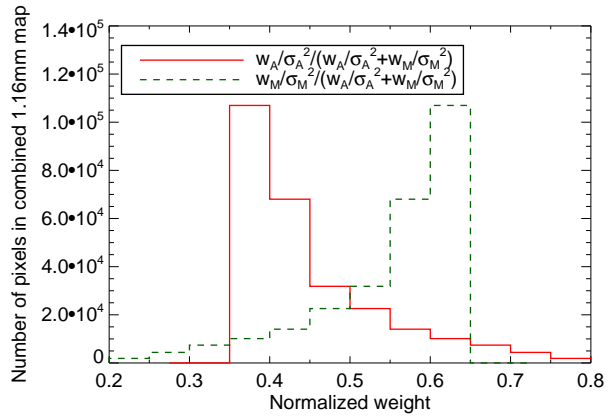}
\caption{Distribution of normalized weights (Eq.~\ref{eqn:coadd})
for pixels with $1\sigma < 1$$\,$mJy
in the combined 1.16$\,$mm map.  The normalized weights applied to the AzTEC and MAMBO
maps for each pixel sum to 1, so the histograms are symmetric about 0.5.
\label{fig:weights}}
\end{figure}

The penalties to pay for these advantages are that the FWHM of the PSF, and the
effective wavelength, vary slightly across the 1.16$\,$mm map.
Alternatively, we could smooth the two individual raw maps to the same PSF resolution, at the expense
of decreased SNR in each pixel. As the weights
(defined as $w_{\mathrm{A}}/\sigma_{\mathrm{A}}^{2}$ and $w_{\mathrm{M}}/\sigma_{\mathrm{M}}^{2}$)
change from pixel to pixel, we average different proportions of AzTEC 1.1$\,$mm and MAMBO 1.2$\,$mm
flux densities.  Fig.~\ref{fig:weights} shows the distributions of
normalized weights (defined in the legend) from Eq.~\ref{eqn:coadd} for pixels with $1\sigma < 1$$\,$mJy in
the combined 1.16$\,$mm map.  The majority of pixels in the combined map are in a small
range of normalized weights ($\sim0.4$ for the AzTEC map, $\sim0.6$ for the MAMBO map);
the variation in FWHM and effective wavelength is small.
The central wavelength
of the combined map is calculated using these normalized weights and the
quoted wavelengths of the two individual maps.
The distribution of stacked flux densities for randomly chosen pixels
in the combined map has zero mean, as expected based on the individual maps
(\S\ref{sec:stacking}).

The area in our combined 1.16$\,$mm map with $1\sigma < 1$$\,$mJy is 0.082$\,$deg$^{2}$.
We use the overlap between this region and the area covered by the
24$\,$$\micron$ sources (0.068$\,$deg$^{2}$) for the stacking analysis.
While we focus on stacking using the combined 1.16$\,$mm map due to its uniform depth (reaching
$1\sigma \sim 0.5$$\,$mJy), we also
compare the stacking results using the SCUBA 850$\,$$\micron$ survey of the GOODS-N region.
The cleaned (of secure detections) 850$\,$$\micron$ map \citep{pope05} has a non-uniform, non-contiguous
0.031$\,$deg$^{2}$ area with $0.5 < 1\sigma < 5$$\,$mJy.
We ensure that both the clean and full SCUBA maps have a mean flux density of 0 mJy in the
area with 24$\,\micron$ sources.

Our terminology is as follows:
\emph{map} refers to a map convolved with its PSF,
except when prefaced with `raw';
\emph{secure detections} are directly detected sources
in the mm map -- that is, non-spurious sources in the AzTEC 1.1$\,$mm and
MAMBO 1.2$\,$mm maps,
and sources with SNR $\ge 3.8$ in the combined 1.16$\,$mm map (see \S\ref{sec:verify} for a
justification of this threshold); hereafter, when we use the word \emph{sources} we mean
sources in an external catalogue that are not detected in the mm maps.  A \emph{cleaned} map has
all secure detections subtracted before convolution with the PSF (\S\ref{sec:multiple_test}),
whereas a \emph{full} map contains the secure detections.

The combined 1.16$\,$mm map is publicly available at
http://www.astro.umass.edu/$\sim$pope/goodsn\_mm/.

\subsection{Verifying the 1.16$\,$mm map}\label{sec:verify}

Detections in the combined 1.16$\,$mm map are found by searching for peaks in the SNR map.
As the SNR threshold is decreased, there is an increasing probability that some detections
are spurious.  \citet{perera08} and \citet{greve08} determine which detections, in their
AzTEC 1.1$\,$mm and MAMBO 1.2$\,$mm maps, are most likely
spurious; most spurious detections have SNR (before deboosting) $< 3.8$, and
only 5 secure detections have SNR (before deboosting) $< 3.8$.
We use this SNR threshold to make our secure detection list for the combined map.
Positions and measured flux densities of secure 1.16$\,$mm detections are given in
Table \ref{table:sources}.

Flux boosting is an important issue for detections at low SNR thresholds,
particularly when the differential counts distribution (d$N/$d$S$) is steep,
so that it is more likely for a faint detection's flux density to scatter up
than for a bright detection's flux density to scatter down.
Flux deboosting is a statistical
correction to the measured flux density of a secure detection \citep{hogg98}.
The deboosting correction relies on a
a simulated map using a model of the differential counts distribution \citep[see][]{coppin05}.
A simulation of the 1.16$\,$mm map is subject to large uncertainties because we
do not have exact knowledge of the PSF, so we choose to deboost the flux densities of secure detections
using an empirical approach.

To verify our method of combining the two maps, we want to compare the
deboosted flux densities of secure detections in the 1.16$\,$mm map with their deboosted
flux densities in the 1.1$\,$mm and 1.2$\,$mm maps.
Our approach to obtain empirically deboosted flux densities is to fit a
function that relates the \emph{deboosted} flux densities of
secure detections in the 1.1$\,$mm and 1.2$\,$mm maps to the \emph{measured} flux densities and
noise values in those maps.  We then use the derived formula to estimate empirically deboosted
flux densities for the secure 1.16$\,$mm detections from the measured 1.16$\,$mm flux densities
and noises.  We find
\begin{equation}
S_{\mathrm{deboosted}} = 1.55\,S_{\mathrm{measured}}^{0.89} - 2.7\,\sigma,
\label{eqn:deboost}
\end{equation}
where $S_{\mathrm{measured}}$ and $\sigma$ are in mJy.
For the secure AzTEC 1.1$\,$mm and MAMBO 1.2$\,$mm detections, the residuals between the deboosted
flux densities from this relation and the deboosted flux densities in \citet{perera08} and
\citet{greve08} have a standard deviation of 0.1$\,$mJy, an error well below the flux density noise values
in all mm maps.  This formula is only valid in the range of SNR covered by the AzTEC and MAMBO
detections, so we do not deboost the flux density of source 1 (a 15$\sigma$ source).
Table \ref{table:sources} lists the deboosted flux densities for the secure
1.16$\,$mm detections using this relation.
For the main purposes of this paper, flux deboosting is not
necessary since we stack the 1.16$\,$mm flux densities of sources we know to exist from other
observations.

\begin{table*}
\begin{minipage}{6in}
\caption{Secure detections in the combined 1.16$\,$mm map (a weighted
average of AzTEC 1.1$\,$mm and MAMBO 1.2$\,$mm maps).\label{table:sources}}
\begin{tabular}{lllllllll}
\hline
Number & RA & Dec & $S_{\mathrm{measured}}$ & $\sigma$ & SNR & $S_{\mathrm{deboosted}}$ &
AzTEC ID & MAMBO ID \\
 &   &     & mJy & mJy      &            & mJy                      & & \\
\hline
1 & 189.299114 & 62.369436 & 10.26 & 0.68 & 15.0 & \ldots & AzGN01 & GN1200.1 \\
2 & 189.137896 & 62.235510 & 5.24 & 0.57 & 9.1 & 5.2 & AzGN03 & GN1200.2 \\
3 & 189.378717 & 62.216051 & 4.51 & 0.55 & 8.2 & 4.4 & AzGN05 & GN1200.4 \\
4 & 189.297686 & 62.224436 & 4.09 & 0.54 & 7.6 & 3.9 & AzGN07 & GN1200.3 \\
5 & 189.132927 & 62.286617 & 4.22 & 0.56 & 7.5 & 4.1 & AzGN02 & GN1200.13 \\
6 & 189.112273 & 62.101043 & 4.81 & 0.67 & 7.2 & 4.5 & AzGN06 & GN1200.5 \\
7 & 188.959560 & 62.178029 & 5.00 & 0.71 & 7.0 & 4.6 & AzGN04 & GN1200.12 \\
8 & 189.308576 & 62.307210 & 3.46 & 0.57 & 6.1 & 3.1 & AzGN26 & GN1200.6 \\
9 & 189.149018 & 62.119408 & 3.35 & 0.61 & 5.5 & 2.9 & AzGN11 & GN1200.14 \\
10 & 189.190353 & 62.244432 & 3.02 & 0.56 & 5.4 & 2.6 & AzGN08 & \ldots \\
11 & 188.973386 & 62.228058 & 3.10 & 0.60 & 5.1 & 2.6 & AzGN13 & GN1200.15 \\
12 & 189.184207 & 62.327207 & 3.03 & 0.59 & 5.1 & 2.5 & AzGN28 & GN1200.9 \\
13 & 189.138377 & 62.105511 & 3.31 & 0.66 & 5.0 & 2.7 & AzGN12 & \ldots \\
14 & 189.213067 & 62.204995 & 2.88 & 0.57 & 5.0 & 2.4 & AzGN14 & GN1200.25 \\
15 & 189.501924 & 62.269772 & 3.26 & 0.66 & 4.9 & 2.7 & AzGN21 & \ldots \\
16 & 189.202112 & 62.351658 & 3.05 & 0.63 & 4.8 & 2.5 & \ldots & \ldots \\
17 & 189.214098 & 62.339995 & 2.88 & 0.60 & 4.8 & 2.3 & \ldots & \ldots \\
18 & 189.068612 & 62.254326 & 2.61 & 0.55 & 4.7 & 2.1 & AzGN16 & \ldots \\
19 & 189.300036 & 62.203880 & 2.59 & 0.55 & 4.7 & 2.1 & \ldots & GN1200.29 \\
20 & 189.114187 & 62.203822 & 2.61 & 0.57 & 4.6 & 2.1 & AzGN10 & \ldots \\
21 & 189.407721 & 62.292688 & 2.62 & 0.58 & 4.5 & 2.1 & AzGN09 & \ldots \\
22 & 189.400013 & 62.184363 & 2.63 & 0.58 & 4.5 & 2.1 & \ldots & GN1200.17 \\
23 & 189.440268 & 62.148758 & 3.84 & 0.85 & 4.5 & 2.8 & \ldots & \ldots \\
24 & 189.035270 & 62.244279 & 2.46 & 0.56 & 4.4 & 1.9 & AzGN24 & \ldots \\
25 & 189.575648 & 62.241841 & 3.56 & 0.82 & 4.3 & 2.6 & \ldots & \ldots \\
26 & 188.951634 & 62.257458 & 2.84 & 0.66 & 4.3 & 2.1 & AzGN15 & \ldots \\
27 & 189.421566 & 62.206005 & 2.41 & 0.57 & 4.3 & 1.9 & AzGN18 & \ldots \\
28 & 188.942743 & 62.192993 & 3.09 & 0.73 & 4.3 & 2.3 & \ldots & \ldots \\
29 & 189.216774 & 62.083885 & 3.74 & 0.88 & 4.2 & 2.6 & AzGN25 & \ldots \\
30 & 188.920762 & 62.242944 & 3.01 & 0.71 & 4.2 & 2.2 & AzGN17 & \ldots \\
31 & 189.323691 & 62.133314 & 2.74 & 0.67 & 4.1 & 2.0 & \ldots & GN1200.23 \\
32 & 189.033574 & 62.148164 & 2.42 & 0.60 & 4.0 & 1.8 & \ldots & GN1200.7 \\
33 & 189.090016 & 62.268797 & 2.23 & 0.56 & 4.0 & 1.7 & \ldots & \ldots \\
34 & 189.143551 & 62.322737 & 2.44 & 0.61 & 4.0 & 1.8 & \ldots & \ldots \\
35 & 189.258342 & 62.214444 & 2.19 & 0.55 & 4.0 & 1.6 & \ldots & \ldots \\
36 & 189.039961 & 62.255953 & 2.21 & 0.56 & 4.0 & 1.6 & \ldots & \ldots \\
37 & 188.916328 & 62.212377 & 2.97 & 0.75 & 4.0 & 2.1 & \ldots & \ldots \\
38 & 189.327507 & 62.231090 & 2.12 & 0.54 & 3.9 & 1.6 & \ldots & \ldots \\
39 & 189.020746 & 62.114810 & 2.71 & 0.70 & 3.9 & 1.9 & AzGN19 & \ldots \\
40 & 189.238057 & 62.279444 & 2.14 & 0.56 & 3.8 & 1.5 & \ldots & \ldots \\
41 & 189.550659 & 62.248008 & 2.78 & 0.73 & 3.8 & 1.9 & \ldots & \ldots \\
\hline
\end{tabular}

Columns: RA and Dec are in decimal degrees, and are reported from the centre
of the pixel with maximum SNR ($S_{\mathrm{measured}}/\sigma$). 
$S_{\mathrm{measured}}$ and $\sigma$ are the measured flux density and noise in the 1.16$\,$mm
map, and $S_{\mathrm{deboosted}}$ is the deboosted flux density calculated with Eq.
\ref{eqn:deboost}.  The AzTEC ID is from \citet{perera08}, the MAMBO ID is
from \citet{greve08}.
\end{minipage}
\end{table*}

\begin{figure*}
\centering
\includegraphics[scale=.44]{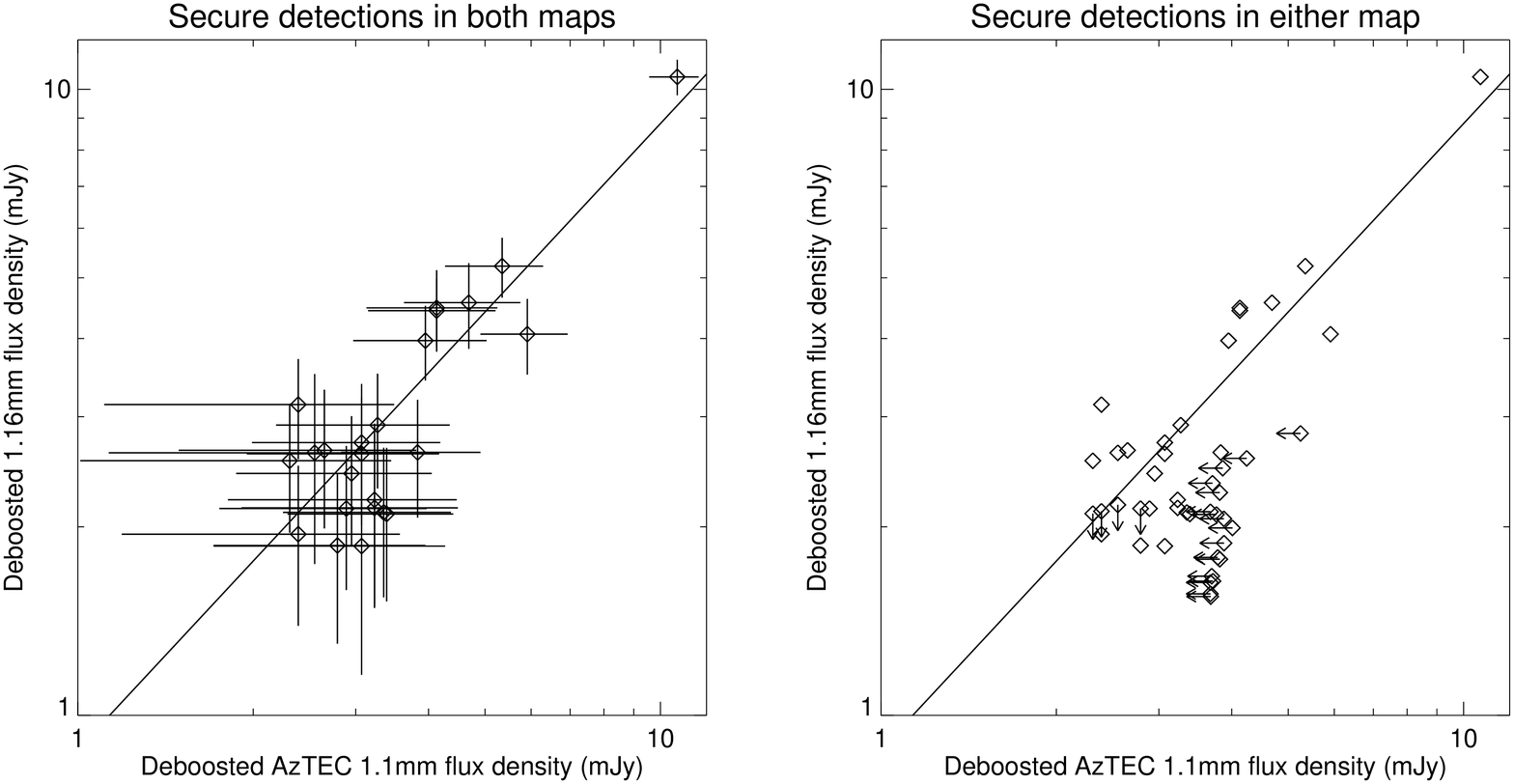}
\caption{
\emph{Left panel}: Empirically deboosted combined 1.16$\,$mm flux
($S_{\mathrm{deboosted}}$) as a function of deboosted AzTEC 1.1$\,$mm flux
($S_{\mathrm{A,deboosted}}$) for detections which are secure in both
maps.  The solid line is the best-fitting line to the deboosted flux densities of
secure detections ($S_{\mathrm{deboosted}} =
0.88S_{\mathrm{A,deboosted}}$).
\emph{Right panel}: A comparison of deboosted flux densities for
secure detections in either map.
If a secure 1.16$\,$mm detection does not coincide with a secure AzTEC 1.1$\,$mm detection, a
$3.8\sigma$ upper limit on the AzTEC 1.1$\,$mm flux density is plotted.  Similarly,
if a secure AzTEC 1.1$\,$mm detection does not coincide with a secure 1.16$\,$mm detection,
a $3.8\sigma$ upper limit on the 1.16$\,$mm flux density is plotted.
\label{fig:deboosted1.1}}
\end{figure*}

Fig. \ref{fig:deboosted1.1} shows the comparison between deboosted
flux densities for secure 1.16$\,$mm and 1.1$\,$mm detections.
The combined 1.16$\,$mm map recovers the majority of
secure detections identified in the AzTEC 1.1$\,$mm map -- the arrows pointing
down show that there are 4 secure detections in the AzTEC map that are not secure detections
in the combined map.

\begin{figure*}
\centering
\includegraphics[scale=.44]{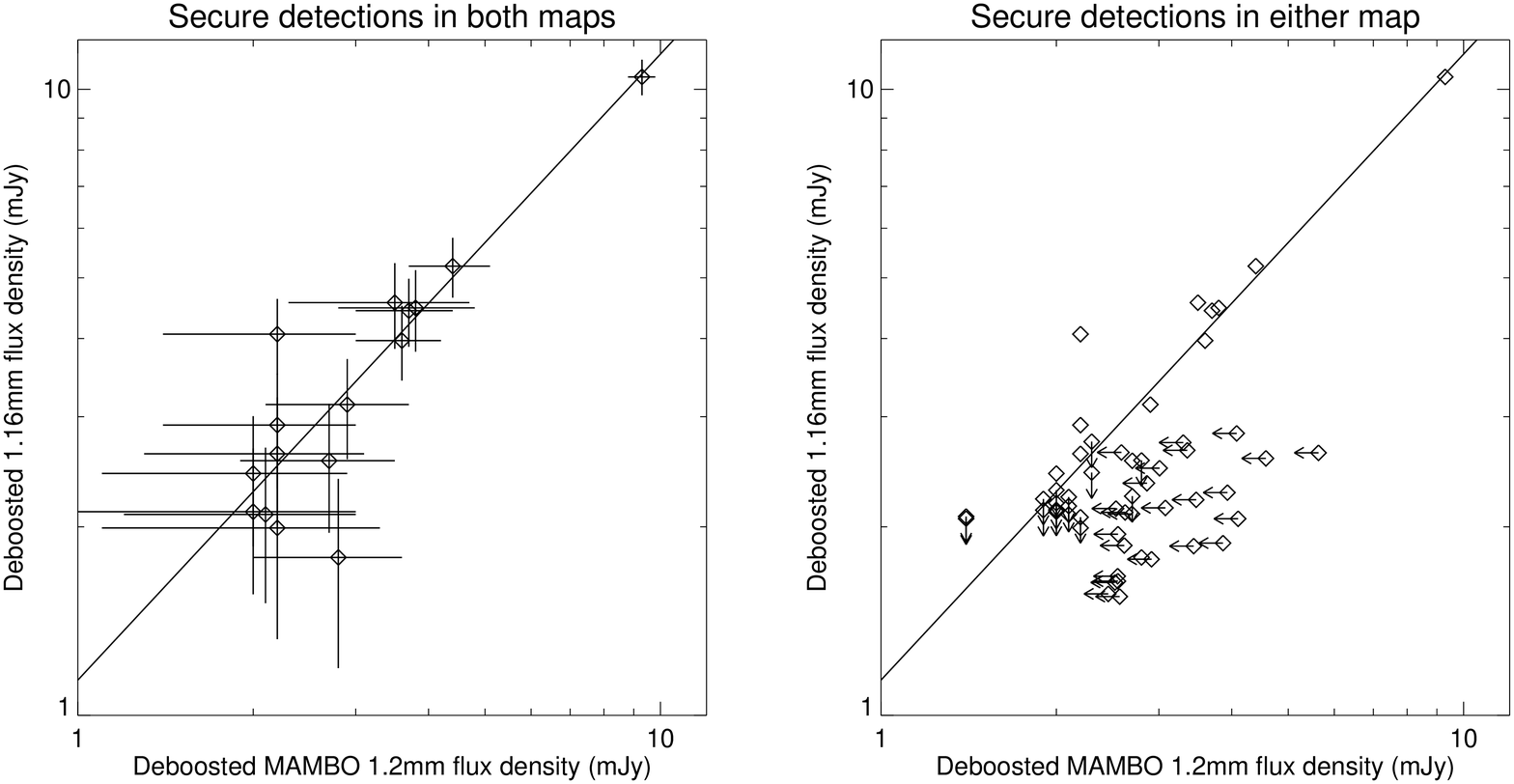}
\caption{\emph{Left panel}: Empirically deboosted combined 1.16$\,$mm flux
($S_{\mathrm{deboosted}}$) as a function of deboosted MAMBO 1.2$\,$mm flux
($S_{\mathrm{M,deboosted}}$) for detections which are secure in both maps. 
The solid line is the best-fitting line to the deboosted flux densities of secure
detections ($S_{\mathrm{deboosted}} = 1.14S_{\mathrm{M,deboosted}}$).
\emph{Right panel}: A comparison of deboosted flux densities for
secure detections in either map.
If a secure 1.16$\,$mm detection does not coincide with a secure MAMBO 1.2$\,$mm detection, a
$3.8\sigma$ upper limit on the MAMBO 1.2$\,$mm flux density is plotted.  Similarly,
if a secure MAMBO 1.2$\,$mm detection does not coincide with a secure 1.16$\,$mm detection,
a $3.8\sigma$ upper limit on the 1.16$\,$mm flux density is plotted.
Based on this figure and Fig.~\ref{fig:deboosted1.1}, we conclude that our method
of combining the AzTEC 1.1$\,$mm and MAMBO 1.2$\,$mm maps is valid.
\label{fig:deboosted1.2}}
\end{figure*}

Fig. \ref{fig:deboosted1.2} shows the comparison between
deboosted flux densities for secure 1.16$\,$mm and 1.2$\,$mm detections.
There are 14 secure detections in the MAMBO 1.2$\,$mm map that
are not coincident with secure detections in the combined 1.16$\,$mm map
(the down arrows in the right panel).  However, the upper
limits to the flux densities in the combined map are
within the scatter about the solid line.

We conclude, based on the comparisons in Figs. \ref{fig:deboosted1.1} and
\ref{fig:deboosted1.2}, that our method of combining the AzTEC 1.1$\,$mm and MAMBO 1.2$\,$mm
maps is effective.  The combined 1.16$\,$mm map has 13 new secure detections (Table
\ref{table:sources}); we do not expect the new detections to be in the individual maps.

\subsection{GOODS-N MIPS 24$\,$$\micron$ redshift catalog}

Galaxies with detected 24$\,$$\micron$ emission compose
the most homogeneous set of dusty galaxies whose mm flux density can be stacked with
significant results.
We use the 24$\,$$\micron$ catalogue from the
GOODS-N \emph{Spitzer}/MIPS survey,
which has a uniform depth of $1\sigma \sim 5$\,$\mu$Jy in the regions of interest (Chary et al. in preparation);
the 24$\,$$\micron$ fluxes are measured at the positions of IRAC sources,
so this catalogue pushes to faint 24$\,$$\micron$ fluxes.
We only stack
$\ge 3\sigma$ 24$\,$$\micron$ sources with $S_{24} > 25$\,$\mu$Jy.  At flux densities above
$50$\,$\mu$Jy, the catalogue is 99 per cent complete; for $25 < S_{24} < 50$\,$\mu$Jy, the
catalogue is 83 per cent complete \citep{magnelli09}.
Completeness corrections to our results are negligible, so we do not apply them.
We exclude sources that
lie in the region of the 1.16$\,$mm map with $1\sigma > 1$$\,$mJy; in this region, the noise is
non-uniform.  The final 24$\,$$\micron$ catalog for stacking has 2484 sources in 0.068 deg$^{2}$.

To decompose the contribution to the mm background from 24$\,$$\micron$ sources
as a function of redshift, we require
either a photometric or spectroscopic redshift for each 24$\,$$\micron$ source.
We start by matching a source with a spectroscopic redshift from the catalogues
of \citet*{barger08} and Stern et al. (in preparation)
to each 24$\,$$\micron$ source.
The match radius, 0.7 arcsec, is chosen by maximizing the number of
unique matches while minimizing the number of multiple matches.  We find
spectroscopic redshifts for 1026 (41 per cent of the) 24$\,$$\micron$ sources.

If no (or multiple) coincident sources with spectroscopic redshifts are found,
we resort to the photometric redshift source catalogue of Brodwin et al. (in preparation)
to find a source match.
Photometric redshifts are constrained with deep $UBVRIzJK$ imaging, and provide redshift
estimates for 872, or 35 per cent, of the 24$\,$$\micron$ sources.
Photometric redshift uncertainties are small compared to our redshift bins, since
we are interested in the contribution to the background from galaxies in large redshift bins.
If no (or multiple)
coincident sources with photometric redshifts are found, we assign the 24$\,$$\micron$ source
to a `redshift unknown' bin in the stacking analysis.  Of the 2484 24$\,$$\micron$ sources,
588 (24 per cent) have no spectroscopic or photometric redshift estimate available.

\section{Stacking analysis}\label{sec:stacking}

Our stacking procedure depends on two fundamental properties
of the (sub)mm maps:
\begin{enumerate}
\renewcommand{\theenumi}{(\arabic{enumi})}
\item \emph{Every detection, and source, is a point source.}  The PSFs are large; in all 3 maps
the full-width-half-maxima (FWHM) are $>$ 10 arcsec.  This property has a number of
implications.  To make low SNR detection-finding easier, the raw maps are
convolved with their PSFs;
the result is a map where each pixel value is the
flux density of a point source at the position of the pixel.
Thus, to stack the millimeter flux densities of sources, we
require only the values of single pixels in the map. 
\item \emph{The means of the maps are 0$\,$mJy.}  These millimeter
observations are taken, filtered, and reduced in such a way that
the sum of all pixel values in the map is zero.  In other words, the most likely value
of a randomly chosen pixel is 0$\,$mJy, a useful statistical property we explore
in \S\ref{sec:multiple}.  However, the large PSF forces us
to carefully consider the effects of having multiple sources clustered in the
area covered by one PSF (also in \S\ref{sec:multiple}).
\end{enumerate}

Stacking is the process of averaging the flux density, at some wavelength (1.16$\,$mm),
of sources detected at another wavelength.
To resolve the (sub)mm background, we want to stack a catalogue
of sources whose emission correlates strongly with 1.16$\,$mm emission, and we want this catalogue
to have a large number of sources.  A catalogue that meets these requirements has galaxies selected on dust
emission at both low and high redshift.  We do not expect a sample of stellar mass selected sources
(for example, at 3.6$\,$$\micron$)
to be efficient at isolating
the galaxies responsible for the mm background, because 3.6$\,$$\micron$
sources are a mix of dusty and non-dusty galaxies.  The MIPS catalogue of
24$\,$$\micron$ sources, though, is selected on dust emission to
high redshift, and there is a known correlation between the flux densities at mid-infrared and
far-infrared wavelengths \citep{chary01}.

The stacking equation we use is similar to Eq.~\ref{eqn:coadd}:
\begin{equation}
S_{\mathrm{bin}} = \frac{\sum_{\mathrm{i}=1}^{N_{\mathrm{bin}}}
\frac{S_{\mathrm{i},1.16}}{\sigma_{\mathrm{i},1.16}^{2}}}{\sum_{\mathrm{i}=1}^{N_{\mathrm{bin}}}
\frac{1}{\sigma_{\mathrm{i},1.16}^{2}}},
\label{eqn:stack}
\end{equation}
where $S_{\mathrm{bin}}$ is the stacked flux density of $N_{\mathrm{bin}}$
sources in a bin of 24$\,$$\micron$ flux density or redshift, and $S_{\mathrm{i},1.16}$ and
$\sigma_{\mathrm{i},1.16}$ are the measured 1.16$\,$mm flux density and noise at the
position of the $i$-th 24$\,$$\micron$ source.  This equation does not include any
constant terms ($w$'s) because the goal of stacking is to get an average flux density for all sources
from a map at one wavelength.  The noise decreases with the inclusion of more sources:
\begin{equation}
\sigma_{\mathrm{bin}} = \frac{1}{\sqrt{\sum_{\mathrm{i}=1}^{N_{\mathrm{bin}}} \frac{1}{\sigma_{\mathrm{i},1.16}^{2}}}}.
\label{eqn:stackrms}
\end{equation}
In a mathematical sense, this equation is only valid when all of the
$\sigma_{\mathrm{i},1.16}$ are independent;
because there are many 24$\,$$\micron$ sources in the area of one PSF,
this requirement is strictly not met.
We fit a Gaussian to the distribution of stacked flux densities for 2484 random pixels,
and the $\sigma$
is the same as the $\sigma_{\mathrm{bin}}$ we calculate
for the 24$\,$$\micron$ sources using Eq. \ref{eqn:stackrms}.
We choose $N_{\mathrm{bin}} \sim 220$ sources when binning by 24$\,$$\micron$ flux density,
and $N_{\mathrm{bin}} \sim 660$ sources when binning
by redshift.  These numbers allow adequate SNR for the stacked flux density in each
bin; the redshift bins are larger because we want a differential contribution from
the sources in each bin of redshift, whereas we want a cumulative contribution from the sources in
each bin of flux density.  The contribution to the 1.16$\,$mm background from each bin is
$N_{\mathrm{bin}}$\,$S_{\mathrm{bin}}/A$, where $A$ is the area.  The overlap
between the 1.16$\,$mm map area with $1\sigma < 1$ mJy and the 24$\,$$\micron$ exposure map
defines $A$ (0.068 deg$^{2}$).

\subsection{The effects of angular clustering on stacking analyses}\label{sec:multiple}

The undetected mm emission from a 24$\,$$\micron$ source covers the
area of the mm PSF, so a natural question to ask is: `What happens to the
stacked mm flux density when there are multiple 24$\,$$\micron$ sources in the area
encompassed by one mm PSF?'  We revisit the fundamental properties of
the mm maps to answer this question.

Consider a randomly distributed population of sources.  We are interested
in the best estimate of the mm flux density of source A, a source with many neighbours.
We remember that 0$\,$mJy is the most likely flux density of a
randomly chosen pixel; an equivalent statement is that the total flux
at the position of A from all of A's randomly distributed neighbors is 0$\,$mJy.
To rephrase qualitatively, there are a few
neighbours with angular separations small enough to contribute
positive flux density at the position of A, but there are many more neighbours with angular
separations that are large enough to contribute negative flux density at the position of A.
\emph{If we have randomly distributed sources in the area covered by the mm PSF,
the true flux densities of the sources are the measured flux densities in the mm map}.  \citet{marsden09}
prove that in the case of randomly distributed sources,
stacking is a measure of the covariance between the stacked catalogue and the (sub)mm map.

Let us also consider a population of sources that is \emph{not} randomly distributed --
a population that is angularly clustered (as we expect the 24$\,$$\micron$
sources to be).  If the clustering is significant at angular separations where
the PSF is positive, and if it is negligible at larger angular separations,
the positive contribution at the position of A from the many sources that
have small angular separations is not cancelled out by the negative contribution
from the sources that have large angular separations.
In this case, the measured flux density of A is higher than the true flux density -- and thus, we
cannot blindly stack multiple sources in the same PSF area.
The stacked flux density of angularly clustered sources near secure detections is overestimated
for the same reason.
The ratio of the measured flux densities to the true flux densities for an ensemble of sources
is a function of the angular clustering strength of the sources, the flux densities at the wavelength
we stack at, and the size of the PSF.  We detail our simulation to compute this ratio
for the 24$\,$$\micron$ sources and the
(sub)mm PSF in \S\ref{sec:multiple_test}.  We further consider the angular clustering of sources
with secure (sub)mm detections; the tests we perform suggest that this angular clustering
is the dominant source of overestimating the stacked flux density.

The aim of the next section is to investigate the impact of
angular clustering on the the stacked (sub)mm flux densities of 24$\,$$\micron$
sources.  Using a similar analysis, \citet{chary10} conclude that clustering
leads to a significant overestimate
of the flux density when stacking on BLAST (sub)mm maps with larger
PSFs than those for the SCUBA 850$\,$$\micron$ and 1.16$\,$mm maps.

The angular clustering of 24$\,$$\micron$ sources is uncertain, though spatial
clustering measurements exist \citep{gilli07}.  The assumption we test is that
this spatial (three dimensional) clustering projects to an angular (two dimensional)
clustering, which may lead to an overestimate of the stacked flux density.

\subsection{Quantifying the effects of angular clustering}\label{sec:multiple_test}

The two tests of our assertion of angular clustering are:
\begin{enumerate}
\renewcommand{\theenumi}{(\arabic{enumi})}
\item An estimate of the ratio of measured flux densities to true flux densities for a simulated map
composed solely of 24$\,$$\micron$ sources. This test quantifies the effect of
angular clustering of 24$\,$$\micron$ sources in the area of one PSF.  Here, \emph{true}
flux density is an input flux density, and \emph{measured} flux density is an output
flux density (after the simulation).
\item A comparison of the resolved background from stacking on a
cleaned map with the resolved background from stacking on a full map.  This test
helps address the effect of angular clustering of 24$\,$$\micron$ sources with secure
(sub)mm detections.
\end{enumerate}
Both tests require a well-characterized PSF: for the first, 
in order to create a realistic simulated map, and for
the second, in order to subtract the secure (sub)mm detections to create a clean map.
The 1.16$\,$mm map does not have a well-characterized
PSF, so we perform the tests
for the \citet{perera08} AzTEC 1.1$\,$mm map, with an area defined by $1\sigma < 1$$\,$mJy
(0.070 deg$^{2}$).
We also run the tests for the SCUBA 850$\,$$\micron$ map, with an area defined by
$1\sigma < 5$$\,$mJy (0.031 deg$^{2}$).

\subsubsection{The first test}

Our first test is a simulation of an AzTEC 1.1$\,$mm map composed exclusively of
24$\,$$\micron$ sources.  Using the relation between 24$\,$$\micron$ flux density
and stacked 1.1$\,$mm flux density (the differential form of Fig. \ref{fig:stack_flux}),
we insert best estimates of the 1.1$\,$mm flux densities at the positions of all the
24$\,$$\micron$ sources.  This process preserves the angular clustering of the real 24$\,$$\micron$
sources.  We then convolve the simulated map with the AzTEC PSF, and remeasure
the 1.1$\,$mm flux densities (by stacking).
The stacked flux density, multiplied by the number of sources, is the \emph{measured}
flux density of the entire sample, while the \emph{true} flux density is the
sum of the inserted flux densities.  The ratio of total measured flux density to total true
flux density is $\sim 1.08$.   Due to angular clustering of multiple
sources within the average PSF,
the stacked 1.1$\,$mm flux density of 24$\,$$\micron$ sources appears to be overestimated
by $\sim 8$ per cent.  Different relations between
24$\,$$\micron$ flux density and 1.1$\,$mm flux density that are physically motivated
(for example, from \citealt{chary01})
produce comparable ratios.  This 8 per cent correction to the stacked
1.1$\,$mm flux density is within the uncertainties (for example, from the relation
between 24$\,$$\micron$ and 1.1$\,$mm flux densities).

An alternative test to the one just presented is an extension of the deblending method
in \citet{greve09} and \citet{kur09}.  Deblending is the simultaneous
solution of a system of $Q$ equations that are
mathematical descriptions of the flux densities of blended, angularly clustered sources
($Q$ is the number of sources to be stacked, see \S5.2 and fig. 5 in \citealt{greve09}).
The result of deblending is a vector of the true source flux densities.
Our extension of the methods in \citet{greve09} and \citet{kur09} generalizes the equations
by not assuming a Gaussian PSF -- which does not have the negative parts that are important
for the data we consider here -- but instead uses the AzTEC PSF for deblending the sources in
the AzTEC map.  Our extension does
not account for 24$\,\micron$ undetected sources that
may affect the stacked 1.1$\,$mm flux of 24$\,\micron$ sources.
This deblending procedure gives
the same answer as our simulations: an 8 per cent overestimation of the stacked 1.1$\,$mm flux density.

\subsubsection{The second test}

Our procedure for cleaning the raw AzTEC 1.1$\,$mm map is: 1) for each
secure 1.1$\,$mm detection, scale the PSF to the deboosted flux density;
2) subtract the scaled PSFs from the raw map;
and 3) convolve the residual map with the PSF.
There are two components to the resolved 1.1$\,$mm background:
the contribution to the background from stacking 24$\,$$\micron$ sources, and
the contribution to the background from the secure 1.1$\,$mm detections cleaned from the
map.  The latter is calculated by summing the deboosted flux densities of all the secure
detections and dividing by the area.

\begin{figure}
\centering
\includegraphics[scale=.44]{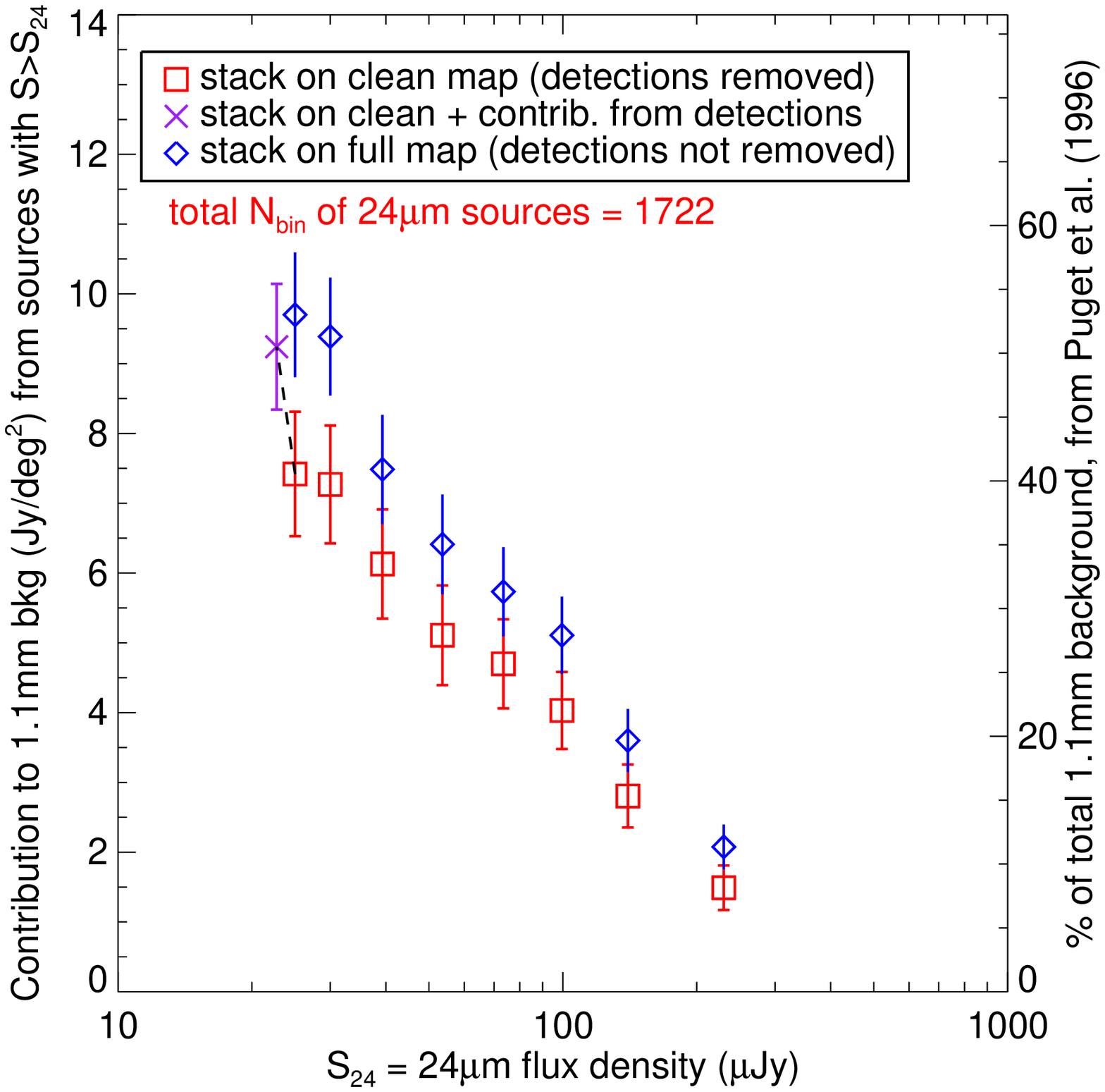}
\caption{The resolved 1.1$\,$mm background
from 24$\,$$\micron$ sources with flux densities $> S_{24}$.
The red squares are a stack on the cleaned map; the
blue triangles are a stack on the full map.  The purple `X' includes
the contribution to the background from the secure 1.1$\,$mm detections, arbitrarily
added to the faintest cumulative flux density bin, after stacking on the cleaned map.
Angular clustering of 24$\,$$\micron$ sources with secure 1.1$\,$mm detections does not appear to cause a 
significant overestimate of the resolved 1.1$\,$mm background.
\label{fig:stack_aztec}}
\end{figure}

We compare the 1.1$\,$mm background resolved from stacking on the full and cleaned
maps in Fig. \ref{fig:stack_aztec} (values in Table \ref{table:multiple}).
A stack of 24$\,$$\micron$ sources on the full map, when compared to a stack on the cleaned
map, does \emph{not} significantly overestimate the resolved 1.1$\,$mm background.

Fig. \ref{fig:stack_aztec}
implies the clustering of 24$\,$$\micron$ sources
with the secure detections in the 1.16$\,$mm map will have a small effect on the stacked flux density,
although we note that the combined 1.16$\,$mm
map does have more secure detections (in a larger area with
$1\sigma < 1$$\,$mJy) than the AzTEC 1.1$\,$mm map.

\begin{figure}
\centering
\includegraphics[scale=.44]{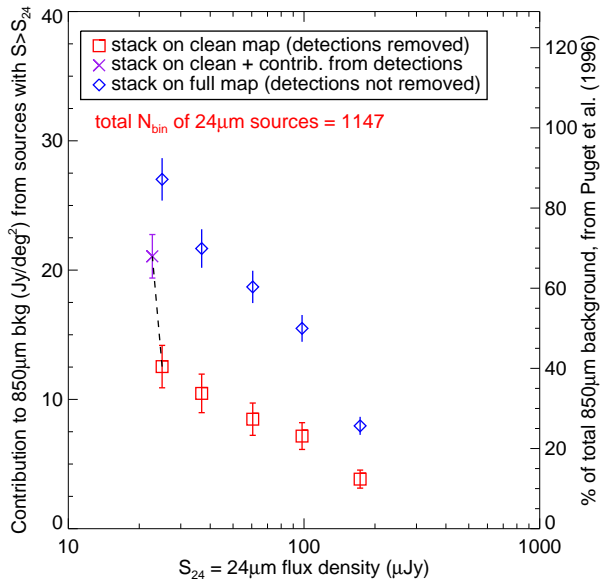}
\caption{The resolved 850$\,$$\micron$ background
from 24$\,$$\micron$ sources with flux densities $> S_{24}$.
The red squares are a stack on the cleaned map; the
blue diamonds are a stack on the full map.  The purple `X' includes
the contribution to the background from the secure 850$\,$$\micron$ detections, arbitrarily
added to the faintest cumulative flux density bin, after stacking on the cleaned map.
In reality, the 24$\,$$\micron$ counterparts to the secure 850$\,$$\micron$ detections have
flux densities ranging from $S_{24}$\,$\sim$\,$20-700$\,$\mu$Jy \citep{pope06}.
We adopt the background values from stacking on the cleaned map.
\label{fig:stack_scuba}}
\end{figure}

The cleaned 850$\,$$\micron$ map is from \citet{pope05}.
We compare the 850$\,$$\micron$ background resolved from stacking on the full and cleaned maps
in Fig. \ref{fig:stack_scuba}.
The blue diamonds (values in Table \ref{table:multiple}) show that a stack of
24$\,$$\micron$ sources on the full 850$\,$$\micron$ map overestimates the resolved submm
background, when compared to a stack on the cleaned map.
We hesitate to attribute the entire difference to angular clustering of 24$\,\micron$ sources with
the secure 850$\,\micron$ detections; the difference is
probably due to many effects:
\begin{enumerate}
\renewcommand{\theenumi}{(\arabic{enumi})}
\item Over-subtraction of the secure 850$\,$$\micron$ detections in making the cleaned map.
Detections are subtracted using measured, rather than deboosted, flux densities.
To estimate the magnitude of this over-subtraction we clean the raw AzTEC 1.1$\,$mm map
using both measured and deboosted flux densities for the
secure 1.1$\,$mm detections, and find a marginal difference in the resolved 1.1$\,$mm background between
the two methods.  The average deboosting correction  -- roughly 30\% of the measured flux subtracted off
(\citealt{perera08}; \citealt{pope06}) --
is similar for both the 850$\,\micron$ and 1.1$\,$mm detections; combined with
the marginal difference in resolved 1.1$\,$mm background, these
suggest that the resolved 850$\,$$\micron$ background is
insensitive to the over-subtraction of the secure 850$\,$$\micron$ detections in the cleaned map.
\item Over-subtraction of the secure 850$\,$$\micron$ detections in regions of
the map close to the confusion limit.  The measured and deboosted flux densities of the detections
in the deepest parts of the 850$\,\micron$ map are not corrected for the contribution from blended sources
below the detection limit.  We compare the background resolved from stacking on the full and cleaned
maps again, this time excluding regions around all detections
with $1\sigma < 1\,$mJy; a large difference in the resolved background remains.
\item Non-uniform noise, which complicates interpretation of the results from
the inverse-variance weighted stacking formula.
\item Different chop throws across the SCUBA map, which complicates the angular separations
where we expect to see negative emission from detections.
\item Angular clustering of the 24$\,\micron$ sources with the secure 850$\,\micron$ detections.
\end{enumerate}
A simulation of the 850$\,\micron$ map, similar to our first test except using \emph{randomly distributed}
sources drawn from a differential counts distribution (d$N/$d$S$) and an idealized SCUBA PSF,
implies that part of the difference may be due to effects other than angular clustering (for example,
effects 1--4).
If this simulation is correct, the stacked 850$\,\micron$ flux density is
\emph{underestimated} when using the cleaned map, and our estimate of the resolved 850$\,\micron$
background is a lower limit.
However, the ratio of stacked 850$\,\micron$ to 1.16$\,$mm flux density
as a function of redshift (using the full 850$\,\micron$ map) requires a model SED with a
higher temperature than 60$\,$K (assuming an emissivity index $\beta$ of 1.5).  We therefore use
the 850$\,\micron$ flux density from stacking on the cleaned map.
With large, uniform maps from SCUBA-2 these issues can be tested and resolved --
until we have such maps, we cannot separate the
effects of angular clustering and non-uniform noise.

\begin{table}
\caption{A comparison of the resolved background at 850$\,$$\micron$ and 1.1$\,$mm
using SCUBA and AzTEC maps (full and cleaned).\label{table:multiple}}
\begin{tabular}{lll}
\hline
Map & 850$\,$$\micron$ bkg & 1.1$\,$mm bkg \\
    &$\,$Jy$\,$deg$^{-2}$ &$\,$Jy$\,$deg$^{-2}$ \\
\hline
Full & $27.0\pm1.6$ & $9.7\pm0.9 $\\
Cleaned & $12.5\pm1.6$ & $7.4\pm0.9$ \\
Cleaned w/detections & $21.1\pm1.7$ & $9.2\pm0.9$ \\
\hline
\end{tabular}
\end{table}

In conclusion, we find that:
\begin{enumerate}
\renewcommand{\theenumi}{(\arabic{enumi})}
\item In the specific
case of the 24$\,$$\micron$ sources and the 1.16$\,$mm map and its PSF,
the effects due to angular clustering are
additional corrections within the statistical uncertainty of the stacked flux density.
\item We cannot separate the effect of angular clustering from the effect of non-uniform noise
in the SCUBA 850$\,\micron$ map.
\end{enumerate}
The results we present in \S\ref{sec:results} use the cleaned 850$\,$$\micron$
map (with the contribution from the secure 850$\,$$\micron$ detections added after stacking)
and the full 1.16$\,$mm map.

\section{Results and discussion}\label{sec:results}

\begin{figure}
\centering
\includegraphics[scale=.44]{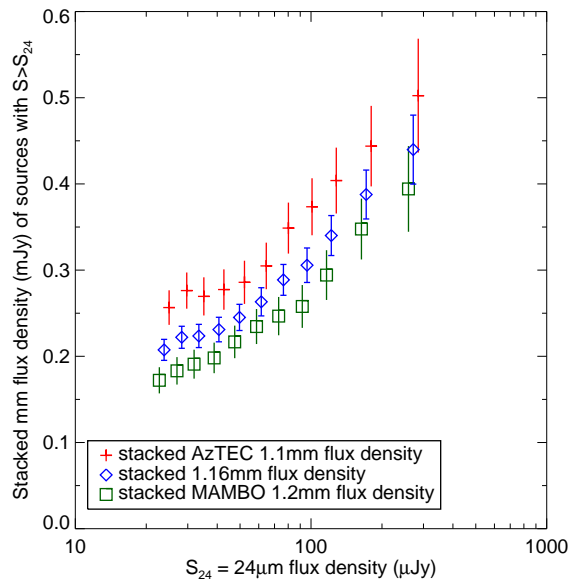}
\caption{Stacked AzTEC 1.1$\,$mm, combined 1.16$\,$mm, and MAMBO 1.2$\,$mm flux densities
from 24$\,$$\micron$ sources with flux densities $> S_{24}$.
The stacked 1.1$\,$mm flux densities are higher than
the stacked 1.2$\,$mm flux densities, as expected for the SED of a typical dusty galaxy.
The 1.16$\,$mm flux density lies between and has smaller errors than the 1.1$\,$mm and
1.2$\,$mm flux densities.
\label{fig:stack_flux}}
\end{figure}

The stacked 1.16$\,$mm flux density as a function of cumulative 24$\,$$\micron$ source flux density is
shown in Fig. \ref{fig:stack_flux}.
This provides another validation of our method of
combining the AzTEC 1.1$\,$mm and MAMBO 1.2$\,$mm maps; the combined 1.16$\,$mm map values (blue diamonds) lie
between the stacked flux densities for the individual maps.  The stack
on the combined 1.16$\,$mm map has smaller errors than the stacks on
the individual maps, as anticipated from Eq.~\ref{eqn:coaddrms}.

\begin{figure}
\centering
\includegraphics[scale=.44]{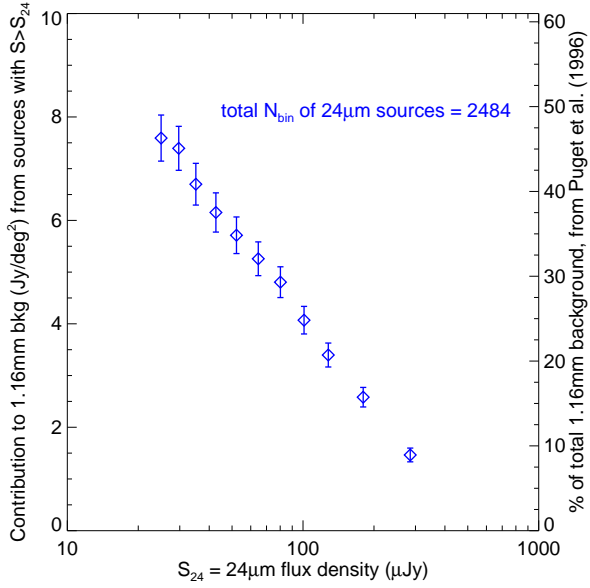}
\caption{Contribution to the 1.16$\,$mm background
from 24$\,$$\micron$ sources with flux densities $> S_{24}$.
\label{fig:stack_bkg}}
\end{figure}

We multiply the stacked flux density (Fig. \ref{fig:stack_flux}) by the number
of 24$\,$$\micron$ sources in the cumulative bin and divide by the area
to get the contribution to the background (Fig. \ref{fig:stack_bkg}).
The overlap between the 1.16$\,$mm map area with $1\sigma < 1$ mJy and the
24$\,$$\micron$ exposure map defines $A$ (0.068 deg$^{2}$).
The blue diamonds show that 24$\,$$\micron$ sources resolve
$7.6\pm0.4$$\,$Jy$\,$deg$^{-2}$ of the 1.16$\,$mm background.

%\begin{table}
%\caption{Resolved 1.16$\,$mm background 
%from 24$\,$$\micron$ sources with flux densities $> S_{24}$ (that
%is, in cumulative bins of flux).
%\label{table:stack}}
%\begin{tabular}{llll}
%\hline
%$S_{24}$ & $S_{\mathrm{bin}}$ & $N_{\mathrm{bin}}$ & 1.16$\,$mm bkg \\
%$\mu$Jy & mJy   &  &$\,$Jy$\,$deg$^{-2}$ \\
%\hline
%285  & $0.440\pm0.040$ & 226 & $1.46\pm.13$ \\
%180  & $0.388\pm0.028$ & 452 & $2.58\pm.19$ \\
%128  & $0.340\pm0.023$ & 678 & $3.40\pm.23$ \\
%101  & $0.306\pm0.020$ & 904 & $4.07\pm.27$ \\
%80.2 & $0.289\pm0.018$ & 1130 & $4.81\pm.30$ \\
%64.6 & $0.263\pm0.016$ & 1356 & $5.26\pm.33$ \\
%52.2 & $0.245\pm0.015$ & 1582 & $5.71\pm.35$ \\
%42.7 & $0.231\pm0.014$ & 1808 & $6.15\pm.38$ \\
%35.1 & $0.224\pm0.013$ & 2034 & $6.70\pm.40$ \\
%29.7 & $0.222\pm0.013$ & 2260 & $7.39\pm.43$ \\
%25.0 & $0.207\pm0.012$ & 2484 & $7.59\pm.45$ \\
%\hline
%\end{tabular}

%Columns: $S_{24}$ is the minimum 24$\,$$\micron$ source
%flux density in the bin.  $S_{\mathrm{bin}}$ is the stacked 1.16$\,$mm flux density of
%$N_{\mathrm{bin}}$ sources with 24$\,$$\micron$ flux densities $> S_{24}$.
%The last column shows the resolved background from
%that cumulative bin.
%\end{table}

The total CIB at (sub)mm wavelengths
is uncertain due to large scale variability of cirrus emission
in the Galaxy that must be subtracted from the observed background, which is
measured using \emph{COBE} maps.
At 1.16$\,$mm, the published estimates
for the total background are 16.4$\,$Jy$\,$deg$^{-2}$ \citep{puget96} and 22.0$\,$Jy$\,$deg$^{-2}$
\citep{fixsen98} (Table \ref{table:bkg}).

\begin{figure*}
\centering
\includegraphics[scale=.45]{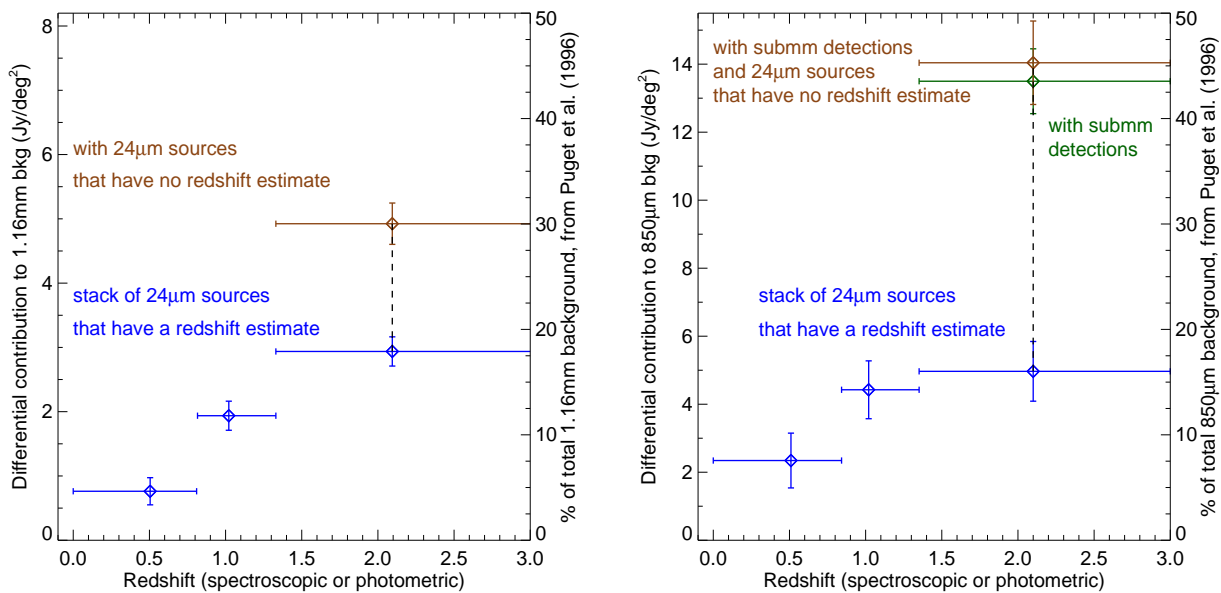}
\caption{\emph{Left panel}:  The (differential) redshift distribution of the resolved
1.16$\,$mm background from 24$\,$$\micron$ sources.
The diamonds are plotted at the average
redshifts of the bins.  The brown diamond contains the contributions from the
24$\,$$\micron$ sources with $z > 1.3$ and the 24$\,$$\micron$ sources
without a redshift estimate.
\emph{Right panel}:  The (differential) redshift distribution of the resolved
850$\,$$\micron$ background from 24$\,$$\micron$ sources.
We use the same redshift bins as in the left panel.
The $y$-axes in both panels show the levels at which the backgrounds
are 50 per cent resolved.  Most of the resolved background at the two
wavelengths comes from galaxies at $z > 1.3$.
\label{fig:stack_all_z}}
\end{figure*}

The left panel in Fig. \ref{fig:stack_all_z} shows
the resolved 1.16$\,$mm background decomposed into redshift bins.
Photometric redshift errors for individual 24$\,$$\micron$ sources
should be negligible in bins of this size.
The highest redshift bin is for all sources with $z > 1.33$,
but we plot it out to $z = 3$ for clarity.
We assume that any 24$\,$$\micron$ sources that fail to match to unique
sources with redshift estimates (either spectroscopic or
photometric) lie at $z > 1.3$, and we add their contribution to
the highest redshift bin.

\begin{table}
\caption{The total background at 4 wavelengths.\label{table:bkg}}
\begin{tabular}{llll}
\hline
Wavelength & Puget 96 & Fixsen 98 & Adopted \\
mm &$\,$Jy$\,$deg$^{-2}$ &$\,$Jy$\,$deg$^{-2}$ &$\,$Jy$\,$deg$^{-2}$ \\
\hline
0.85 & 31 & 44$^{+5}_{-8}$ & $40\pm9$ \\
1.1 & 18.3 & 24.8$^{+1.7}_{-4.0}$ & \ldots \\
1.16 & 16.4 & 22.0$^{+1.4}_{-3.4}$ & $19.9\pm3.5$ \\
1.2 & 15.4 & 20.4$^{+1.1}_{-3.0}$ & \ldots \\
\hline
\end{tabular}
\end{table}

The 1.16$\,$mm background is not fully resolved by 24$\,$$\micron$ sources with
$S_{\mathrm{24}} > 25\,\mu$Jy;
most of the portion that is resolved comes from galaxies at high redshift ($z > 1.3$).
We repeat our stacking analysis on the cleaned 850$\,$$\micron$ map to investigate
the differences in the resolved portions of the background at 850$\,$$\micron$ and 1.16$\,$mm.

We use the same redshift bins as in the 1.16$\,$mm analysis (the right panel
in Fig. \ref{fig:stack_all_z}).
At 850$\,$$\micron$, the values for the total
background are 31$\,$Jy$\,$deg$^{-2}$ \citep{puget96} and 44$\,$Jy$\,$deg$^{-2}$
\citep{fixsen98} (Table \ref{table:bkg}).
The contribution from the 
secure 850$\,$$\micron$ detections is added to the contribution derived from stacking the
24$\,\micron$ sources on the cleaned map; all secure
850$\,$$\micron$ detections have 24$\,$$\micron$ counterparts, and we assume
for simplicity that the detections lie at $z > 1.3$.  This assumption is reasonable,
since only 4 of the 33 detections appear to lie at $z < 1.3$ \citep{pope06}, and
these 4 account for $< 5$ per cent of the contribution from the detections.

Our analysis does not definitively provide the redshift origins of the total
850$\,$$\micron$ background, since it is not completely resolved by 24$\,$$\micron$
sources.  The results suggest that a large fraction
of the resolved 850$\,$$\micron$ background originates in galaxies at $z > 1.3$.
\citet{wang06} perform a stacking analysis and conclude
that more than half of the background at 850$\,$$\micron$ comes from galaxies
at low redshifts ($z < 1.5$).  Our methodology differs from that of
\citet{wang06}: they stack a near infrared ($H$ + 3.6$\,$$\micron$) sample
on a full map with the 850$\,$$\micron$ detections.

\begin{table*}
\begin{minipage}{6.9in}
\caption{The redshift distribution of the resolved background at 1.16$\,$mm
and 850$\,$$\micron$ from 24$\,$$\micron$ sources.\label{table:stack_z}}
\begin{scriptsize}
\begin{tabular}{lllllllll}

\hline
$z$ & $S_{\mathrm{bin},1.16}$ &
$N_{\mathrm{bin},1.16}$ &  per cent w/spec-$z$ & 1.16$\,$mm bkg &
$S_{\mathrm{bin},850}$ & $N_{\mathrm{bin},850}$ &
850$\,$$\micron$ bkg & 850/1.16 \\
    & mJy                &  & &$\,$Jy$\,$deg$^{-2}$ & mJy &  &$\,$Jy$\,$deg$^{-2}$ & \\
\hline
0 -- 0.82 & $0.090\pm0.025$ & 576 & 75 & $0.76\pm0.21$ & $0.237\pm0.081$ & 
304 & $2.34\pm0.81$ & $3.1\pm1.4$ \\
0.82 -- 1.33 & $0.199\pm0.023$ & 660 & 64 & $1.94\pm0.23$ & $0.402\pm0.077$ &
338 & $4.42\pm0.85$ & $2.3\pm0.5$ \\
$>1.33$ & $0.302\pm0.023$ & 660 & 26 & $2.94\pm0.23$ & $0.492\pm0.087$ & 
310 & $4.97\pm0.88$ & $1.7\pm0.3$ \\
%$>1.33$+no z & $0.268\pm0.017$ & 1248 & \ldots & $4.92\pm.32$ & $0.335\pm0.071$ &
%505 & $5.51\pm1.17$ & $1.2\pm.3$ \\
\hline
\multicolumn{9}{c}{with 850$\,$$\micron$ detections added to highest $z$ bin} \\
\hline
$>1.33$ & \ldots & 660 & \ldots & $2.94\pm0.23$ & \ldots &
343 & $13.50\pm0.95$ & $4.6\pm0.5$ \\
\hline
\multicolumn{9}{c}{with `redshift unknown' added to highest $z$ bin} \\
\hline
$>1.33$ & \ldots & 1248 & \ldots & $4.92\pm0.32$ & \ldots &
538 & $14.04\pm1.23$ & $2.9\pm0.3$ \\
\hline
\end{tabular}
\end{scriptsize}

Columns: $S_{\mathrm{bin},1.16}$ is the stacked 1.16$\,$mm flux density of
$N_{\mathrm{bin},1.16}$ sources;
$S_{\mathrm{bin},850}$ is the stacked 850$\,$$\micron$ flux density of
$N_{\mathrm{bin},850}$ sources.  Column 4 is the percentage
of the $N_{\mathrm{bin},1.16}$ sources that have a redshift
determined spectroscopically.  Columns
5 and 8 are the resolved background in each bin.  Column
9 is the resolved 850$\,$$\micron$ background divided by the
resolved 1.16$\,$mm background.
\end{minipage}
\end{table*}

We show that the background at 850$\,$$\micron$ and 1.16$\,$mm is only partially
resolved.  Can we provide any constraints on the redshifts of the galaxies that contribute
to the remainder of the 1.16$\,$mm background?

\begin{figure}
\centering
\includegraphics[scale=.45]{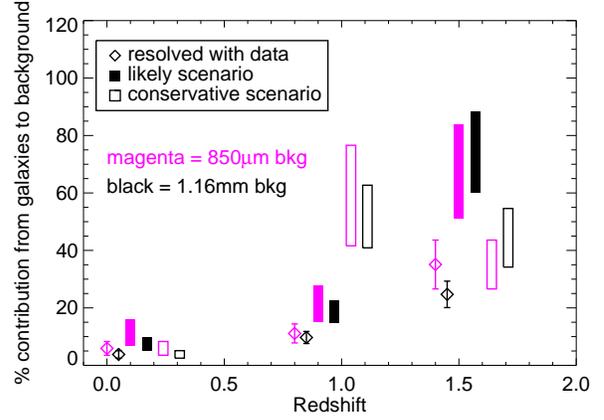}
\caption{The redshift origins of the background at 850$\,$$\micron$ and 1.16$\,$mm
under various scenarios (see \S\ref{sec:results} for details).  The different plotting styles indicate
different scenarios; all magenta points/bars are for the 850$\,$$\micron$ background,
while all black points/bars are for the 1.16$\,$mm background.  Points/bars are offset within the redshift
bins for clarity.  In what we deem
the most likely scenario, 60--88 per cent of the 1.16$\,$mm background comes from galaxies at
$z > 1.3$.
\label{fig:percent_z}}
\end{figure}

There are two often used estimates for the total background at these two wavelengths.
We adopt the average of the range allowed by the two estimates:
$19.9\pm3.5$$\,$Jy$\,$deg$^{-2}$ at 1.16$\,$mm, and $40\pm9$$\,$Jy$\,$deg$^{-2}$
at 850$\,\micron$ (Table \ref{table:bkg}).  If we assume that the galaxies
responsible for the remaining unresolved 850$\,$$\micron$ background are distributed to
maintain the redshift distribution of the galaxies contributing to the resolved
background, then the final decomposition of the 850$\,$$\micron$ background is
[$z \sim 0.4$, $z \sim 1$, $z > 1.3$] = [$4.5\pm1.6$$\,$Jy$\,$deg$^{-2}$,
$8.5\pm1.6$$\,$Jy$\,$deg$^{-2}$, $27\pm2$$\,$Jy$\,$deg$^{-2}$].  The errors maintain
the SNR of the redshift bins of the resolved background.
We also assume that the ratios of the resolved 850$\,$$\micron$ to 1.16$\,$mm background
as a function of redshift (last column of Table \ref{table:stack_z}) hold for the total 850$\,$$\micron$ background;
we thus convert each contribution to the 850$\,$$\micron$ background into an
estimate of the contribution to the 1.16$\,$mm background.
The decomposition of the 1.16$\,$mm background is thus
4.5/3.1$\,$+$\,$8.5/2.3$\,$+$\,$27/2.9$\,$=$\,$$14.4\pm0.85$$\,$Jy$\,$deg$^{-2}$.
The rest of the 1.16$\,$mm background, which is
19.9$\,$--$\,$14.4$\,$=$\,$$5.4\pm0.85$$\,$Jy$\,$deg$^{-2}$, presumably comes from galaxies at
$z > 1.3$, where the observed submm to mm flux density ratio is lower than the
values we use (see, for example, fig. 13 in \citealt{greve04}).
The sum of all contributions from galaxies at $z > 1.3$ is $14.8\pm1.1$$\,$Jy$\,$deg$^{-2}$,
or $74\pm14$ per cent of the total 1.16$\,$mm background.  This likely scenario for the unresolved background is shown with
filled bars in Fig. \ref{fig:percent_z}.

Although we cannot quantify the probability that the unresolved 850$\,$$\micron$ background is
distributed as the resolved background, we are able to derive a lower limit to the amount
of the total 1.16$\,$mm background that comes from galaxies at $z > 1.3$.
In a conservative scenario, all of the
remaining unresolved 850$\,$$\micron$ background comes from galaxies at
$z < 1.3$.  Assuming the ratio of 2.3 at $z \sim 1$ holds for the total background,
an additional contribution of 40$\,$--$\,$2.3$\,$--$\,$4.4$\,$--$\,$14$\,$=$\,$19.3$\,$Jy$\,$deg$^{-2}$ at 850$\,$$\micron$
corresponds to an additional
contribution of 8.4$\,$Jy$\,$deg$^{-2}$ at 1.16$\,$mm.  If the unresolved 850$\,$$\micron$ background
is produced only by $z < 1.3$ galaxies, the contribution to the 1.16$\,$mm background is
0.8$\,$+$\,$10.3$\,$+$\,$4.9$\,$=$\,$$16\pm1.3$ $\,$Jy$\,$deg$^{-2}$.
Again, the remaining 19.9$\,$--$\,$16$\,$=$\,$$3.9\pm1.3$ of the 1.16$\,$mm background comes from galaxies at $z > 1.3$.
At minimum, $44\pm10$ per cent of the total 1.16$\,$mm background comes from galaxies at $z > 1.3$.
This conservative scenario is illustrated with unfilled bars in Fig. \ref{fig:percent_z}.

An alternate explanation to both scenarios is that all the unresolved background
comes from a population of low redshift galaxies with very cold dust and no
warm dust (that is, a population
of galaxies with a disproportionate amount of large dust grains relative to
small dust grains).  Our decomposition of the (sub)mm background depends
on selecting dusty galaxies at 24$\,$$\micron$ -- the selection could miss galaxies
with little or no warm dust.  Galaxies with an excess of cold dust need dust temperatures in the realm of
$\sim10\,$K at $z \sim 1$, and lower temperatures at lower redshifts,
to account for the ratio of unresolved 850$\,\micron$ to 1.1$\,$mm background; large numbers
of galaxies are unlikely to have these extreme dust temperatures.

In this paper, we use observational constraints on the fraction of the (sub)mm background that is resolved
to hypothesize that 60--88 per cent of the 1.16$\,$mm background comes from high redshift
galaxies.  In order to resolve the total 1.16$\,$mm background and
provide direct constraints on the redshifts of the galaxies, we need improvements in both
the catalogue to be stacked and the mm map.
Any stacking catalogue must be deep and homogeneously selected across a
large redshift range.  The GOODS-N survey at 100 $\micron$ with \emph{Herschel}
will reach similar (total infrared luminosity) depths as the deepest surveys at
24$\,$$\micron$ with \emph{Spitzer}; furthermore,
the flux density from 100$\,$$\micron$ sources should correlate more tightly with mm flux density
than does the flux density from 24$\,$$\micron$ sources (dust emitting at 100$\,$$\micron$
is a better tracer of the dust emitting at 1$\,$mm).  Much deeper radio catalogues than
currently exist for stacking, using EVLA and ALMA, are also promising.
Alternatively, future large dish (sub)mm telescopes,
such as the Large Millimeter Telescope, will provide maps in which the bulk of the galaxies that
contribute to the cosmic millimeter background are individually detected.  Models
presented in \citet{chary10} predict that 60 per cent of the 1.2$\,$mm background comes from galaxies with 1.2$\,$mm flux densities
larger than 0.06$\,$mJy (30 times deeper than the combined map).

\section{Conclusions}\label{sec:conclusions}

\begin{enumerate}
\renewcommand{\theenumi}{(\arabic{enumi})}
\item We create a deep ($\sigma \sim 0.5$$\,$mJy) 1.16$\,$mm map by averaging the AzTEC 1.1$\,$mm
and MAMBO 1.2$\,$mm maps in the GOODS-N region.  We verify the properties of this
map by examining both the deboosted flux densities of the 41 secure detections and the
stacked flux density of 24$\,$$\micron$ sources.  Of the 41 secure detections, 13 are new.

\item We test the effects of angular clustering of 24$\,$$\micron$ sources
on the stacked (sub)mm flux density.  While clustering does not seem to lead to a significant overestimate
of the stacked 1.16$\,$mm flux density, it may be responsible for part of the overestimate of
the stacked 850$\,$$\micron$ flux density.

\item 24$\,$$\micron$ sources resolve 7.6$\,$Jy$\,$deg$^{-2}$ (31--45 per cent) of the 1.16$\,$mm
background; 3$\,$Jy$\,$deg$^{-2}$ comes from galaxies at $z > 1.3$.  24$\,$$\micron$ sources
resolve 12.3$\,$Jy$\,$deg$^{-2}$ (23--39 per cent) of the 850$\,$$\micron$ background, and the submillimeter
detections contribute an additional 16--26 per cent; 14$\,$Jy$\,$deg$^{-2}$ of the 850$\,\micron$
background comes from galaxies at $z > 1.3$.

\item Using the ratio of the resolved 850$\,$$\micron$ background to the
resolved 1.16$\,$mm background, we propose that 60--88 per cent of the cosmic
millimeter background comes from high redshift ($z > 1.3$) galaxies.  In the most conservative scenario,
34--55 per cent of the 1.16$\,$mm background comes from galaxies at $z > 1.3$.

\end{enumerate}

We hope to directly detect the majority of the galaxies contributing to the millimeter background with
future surveys using large telescopes (for example, the LMT).  Deeper catalogues for stacking, at radio
and far infrared wavelengths, are needed to fully resolve the mm background.
Future studies will also need to assess the effects of angular clustering.

\section*{Acknowledgments}

We thank the referee for their helpful comments.
This work is based on observations made with the
\emph{Spitzer} Space Telescope, which is operated by the Jet Propulsion
Laboratory, California Institute of Technology under a contract with NASA.
Support for this work was provided by NASA through an award issued by
JPL/Caltech.  AP acknowledges support provided by NASA through the
\emph{Spitzer} Space Telescope Fellowship Program, through a contract
issued by the Jet Propulsion Laboratory, California Institute of Technology
under a contract with NASA.
The Dark Cosmology Centre is funded by the DNRF.
TRG acknowledges support from IDA.
Support for MB was provided by the W.~M.~Keck Foundation.
KC acknowledges the UK Science and Technology Facilities Council (STFC) for a fellowship.

\end{document}